\documentclass[preprint,12pt]{elsarticle}
\usepackage{amsmath,amsfonts}
\usepackage{algorithmic}
\usepackage{graphicx}
\usepackage{textcomp}
\usepackage{xcolor}
\usepackage{xspace}
\usepackage{hyperref}
\usepackage{tabularray}
\usepackage{verbatim}
\usepackage{booktabs}
\usepackage{multirow}
\usepackage{multicol}
\usepackage{amsmath}
\usepackage{lipsum}
\usepackage{array}
\usepackage{makecell}
\usepackage[acronym,nohypertypes={acronym}]{glossaries}
\usepackage{orcidlink}
\usepackage{amssymb}
\usepackage{pifont}%
\usepackage{array}
\usepackage{ragged2e}
\usepackage{adjustbox}
\usepackage[numbers]{natbib}

\newcommand{\etal}[0]{\emph{et al.}\xspace}
\newcommand\ie{\emph{i.e.},\xspace}
\newcommand\eg{\emph{e.g.},\xspace}
\newcommand\etc{\emph{etc.}}
\newcommand{\citesec}[1]{Section~\ref{sec:#1}}

\newcommand{\ghc}[0]{\textsc{GitHub Copilot}}

\newcommand{\cmark}{\ding{51}}%
\newcommand{\xmark}{\ding{55}}%

\newacronym{llm}{LLM}{\textit{Large Language Model}}
\newacronym{tgi}{TGI}{\textsc{Text Generation Inference}}
\newacronym{ide}{IDE}{\emph{Integrated Development Environment}}
\newacronym{ai}{AI}{Artifical Intelligence}
\newacronym{cs}{CS}{Computer Science}
\newacronym{ghc}{GHC}{\textsc{GitHub Copilot}}

\usepackage{changepage}
\usepackage{framed}
\makeatletter
 {\par\unskip\endMakeFramed}
\makeatother

\definecolor{formalshade}{rgb}{0.93,0.93,0.93}
\definecolor{darkblue}{rgb}{0.2, 0.2, 0.2}

\newenvironment{formal}{%
  \def\FrameCommand{%
    \hspace{1pt}%
    {\color{darkblue}\vrule width 2pt}%
    {\color{formalshade}\vrule width 4pt}%
    \colorbox{formalshade}%
  }%
  \MakeFramed{\advance\hsize-\width\FrameRestore}%
  \noindent\hspace{-1pt}%
  \begin{adjustwidth}{}{7pt}%
  \vspace{2pt}\vspace{2pt}%
}
{%
  \vspace{3pt}\end{adjustwidth}\endMakeFramed%
}

\newcounter{resultcounter}

\begin{document}
\begin{frontmatter}
\title{\textsc{Green}\,My\,LLM: Studying the key~factors affecting the~energy~consumption of code~assistants}

\author[univ]{Tristan Coignion} %
\ead{tristan.coignion@inria.fr}
\author[univ]{Cl\'ement Quinton}
\author[univ]{Romain Rouvoy}

\affiliation[univ]{organization={Univ.\,Lille, CNRS, Inria, Centrale Lille, UMR 9189 CRIStAL, F-59000 Lille},
            country={France},}

\begin{abstract}
In recent years, \glspl{llm} have significantly improved in generating high-quality code, enabling their integration into developers' \glspl{ide} as code assistants.
These assistants, such as \textsc{GitHub Copilot}, deliver real-time code suggestions and can greatly enhance developers' productivity. 
However, the environmental impact of these tools, in particular their energy consumption, remains a key concern. 
This paper investigates the energy consumption of \gls{llm}-based code assistants by simulating developer interactions with \textsc{GitHub Copilot} and analyzing various configuration factors. 
We collected a dataset of development traces from $20$ developers and conducted extensive software project development simulations to measure energy usage under different scenarios.

Our findings reveal that the energy consumption and performance of code assistants are influenced by various factors, such as the number of concurrent developers, model size, quantization methods, and the use of streaming. 
Notably, a substantial portion of generation requests made by \textsc{GitHub Copilot} is either canceled or rejected by developers, indicating a potential area for reducing wasted computations. 
Based on these findings, we share actionable insights into optimizing configurations for different use cases, demonstrating that careful adjustments can lead to significant energy savings.
\end{abstract}


\end{frontmatter}

\glsresetall
\section{Introduction}\label{sec:intro}
In recent years, \glspl{llm} for code have significantly become better at generating code, facilitating their seamless integration into developers' \glspl{ide} as code assistants. 
Code assistants offer auto-completion suggestions that developers can either accept or reject. 
The generation process is typically initiated automatically after a brief pause in typing, but can also be manually triggered via a command or keyboard shortcut.

Numerous code assistants, like \textsc{GitHub Copilot}, \textsc{Tabnine}, and \textsc{CodeWhisperer}, including open-source options---like \textsc{Tabby} and \textsc{Cody}---offer \gls{ide} extensions and manage inference servers for code suggestions. 

However, the energy consumption of software has gained prominence as a significant environmental and societal concern. 
In the context of \gls{ai}, Green\,AI has been defined by Schwartz \etal~\cite{schwartzGreenAIPaper2020} as research that yields novel results while considering the computational cost, making practitioners ideally reduce resources spent. 
Specifically to Green\,AI applied to \glspl{llm}, studies have observed the environmental impact of training and using such \glspl{llm}. 
For instance, Samsi~\etal benchmarked the energy consumption of \gls{llm} inference and were able to estimate the energy of a single response from an \gls{llm}~\cite{samsiWordsWattsBenchmarking2023}. 
Other works focused more on the impact of training the model, such as the carbon footprint of the \textsc{Bloom} model estimated by Luccioni~\etal~\cite{luccioniPowerHungryProcessing2023}.
However, the evaluation of \glspl{llm}' energy consumption remains challenging when assessing \glspl{llm} dedicated to specific purposes. 
In the context of \glspl{llm} for code, existing studies have only focused on the impact of the generated code~\cite{vartziotisLearnCodeSustainably2024, coignion:hal-04525620}.

In this paper, we aim to determine how much energy an average developer consumes when using a code assistant similar to \textsc{GitHub Copilot} and how to reduce it. 
To the best of our knowledge, our work is one of the first to study the energy consumption of \glspl{llm} in code assistants. 
In particular, we wish to deliver actionable insights into the energy consumption of the code assistant from the perspective of both the service provider (\eg \textsc{GitHub Copilot}) and the end-user interacting with the code assistant. 
All the more in the context of code assistants, like \textsc{GitHub Copilot}, the end-user knows little about the internal workings and impacts of the service:
as the \gls{llm} inference is executed remotely in the cloud, it is difficult for the developer to perceive the computing impact of using a code assistant.
Thus, we aim to answer the following research questions:
\begin{description}
    \item[\textbf{RQ1}:]  \textit{What is the impact of certain factors 
    on the energy consumption and performance of code assistants?}
    Specifically, we study the impacts of the number of concurrent developers using the assistant, the streaming and manual triggering of the requests, the model and its quantization, the maximum number of concurrent requests, and the number of GPUs.

    \item[\textbf{RQ2}:] \textit{How many generation requests made by \textsc{GitHub Copilot} are useful?} 
    Knowing how many generations are useful to the developer could allow future works to improve the efficiency of code assistants.
    
    \item[\textbf{RQ3}:] \textit{How much energy does a developer consume when using a code assistant similar to \textsc{GitHub Copilot} under different scenarios and objectives?}
    We aim to get a broad look at the potential impacts of a code assistant by leveraging the knowledge about the various factors we gathered when answering RQ1.
\end{description}

We chose to study \textsc{GitHub Copilot} specifically for three reasons: \emph{(i)} it is the most used code assistant according to the 2023 StackOverflow developer survey,\footnote{\url{https://survey.stackoverflow.co/2023/\#section-most-popular-technologies-ai-developer-tools}} \emph{(ii)} the inference server provided by \textsc{GitHub Copilot} exhibits low latency and correct quality~\cite{doderleinPilotingCopilotCodex2022}, which we may not be able to reproduce with our own inference server, and \emph{(iii)} \textsc{GitHub Copilot} provides many mechanisms making it one of the state-of-the-art code assistant, such as preventing some generations requests that may not be useful, caching the results of the previous generations requests, canceling the previous request when a new one is sent, and building of prompts that take into accounts multiple files. 
We are aware of the existence of these mechanisms thanks to the \textsc{Copilot Explorer} project~\cite{Copilotexplorera}, while evidence of similar mechanisms in other code assistants is unclear, and verification is time-consuming.

In this paper, we share the following contributions:
\begin{itemize}
    \item[-] We provide the first dataset of development traces from developers using \textsc{GitHub Copilot}, which allows the simulation of developers using a code assistant on a real inference server.
    \item[-] We study the impact of some configuration options of the inference server and the code assistant on its energy consumption and performance. 
    \item[-] We analyze the ratio of useful generation requests from \textsc{GitHub Copilot}.
    \item[-] We estimate how much energy a developer would consume when developing under different configurations of the inference server.
\end{itemize}

\textbf{Main findings \& implications.}
Our results reveal that a considerable portion of energy is wasted due to suggestions being cancelled or not wanted by the users. 
Thus, manually triggering the generation of the code assistant can significantly reduce its energy consumption, by minimizing unnecessary requests. 
Additionally, the number of concurrent developers on a single inference server greatly affects the energy consumption of a single developer, as having more concurrent developers better uses the resources of the server. 
Providers should aim to maximize the number of developers per server or share inference servers in order to maximize efficiency.
Lastly, the configuration of the inference server of the code assistant plays a significant role in its energy consumption. 
For instance, smaller and models tends to use less electricity, reducing the number of GPUs decreases the energy consumption at the cost of a higher latency, and some quantization techniques can positively both the energy consumption and latency of the code assistant.

\textbf{Outline.}
From \autoref{sec:Dataset collection} to \autoref{sec:data_analysis}, we describe how we collected our dataset, performed our experiments, and explain the methodology we followed to analyze the results obtained.
We report in \autoref{sec:results} on the results of our evaluation and provide a critical discussion in \autoref{sec:discussion}. 
In \autoref{sec:limits}, we discuss the limitations of our study. 
Finally, \autoref{sec:rw} presents related works, and \autoref{sec:conclusion} concludes the paper.

\section{Code Assistant Dataset}\label{sec:Dataset collection}
To investigate our research questions, it was imperative to measure the energy consumption of a code assistant in a realistic usage setting. 
To that end, we designed an experiment with participants using \textsc{GitHub Copilot} to gather a dataset of development traces, enabling us to simulate developers using a code assistant on an inference server under our control. 
This approach was necessitated by our inability to access \textsc{GitHub Copilot}'s inference server and our objective to measure its power consumption.
Moreover, conducting this experiment in two distinct phases---(i) having participants use a code assistant to develop a small application followed by (ii) exploring the energy consumption through traces replay---allowed us to gain more freedom when it came to controlling the configuration of the code assistant and facilitated the reproducibility of our experiment. 
Hereafter, we refer to the traces dataset we collected as \textsc{AssistantTraces}.

\subsection{Involved participants}
We recruited $20$ volunteers among \gls{cs} students and \gls{cs} professionals, using mailing lists and word of mouth. 
All of the participants were proficient in Java programming.
$13$ participants were between $18$ and $25$ years old, $6$ were between $26$ and $35$ years old, and $1$ was between $36$ and $45$ years old.
$13$ of them were students in \gls{cs} ($12$ in a master's degree and $1$ in a bachelor's degree), and $7$ of them were \gls{cs} professionals.
$1$ participant did not know \textsc{GitHub Copilot} before the experiment, $7$ participants knew \textsc{GitHub Copilot} but never used it, $4$ already used it a little, and $8$ used it regularly. 
All of the participants were familiar with \textsc{VSCode}.
The participants were compensated for their time with 50€ each. 
This experiment was approved by our institution's Ethical Board.

Before the experiment, the $20$ participants filled out a survey with their age, development experience, and familiarity with \textsc{GitHub Copilot}. 
Following the experiment, $19$ out of the $20$ participants agreed to complete a subsequent survey concerning their experience and feelings during the experiment. 
This last survey was used to report ideas for future research.

\subsection{Assigned task}
The task given to the participants consisted of developing a CLI Connect\,4 game\footnote{\url{https://en.wikipedia.org/wiki/Connect_Four}} in Java in one hour using \textsc{VScode} and \textsc{GitHub Copilot}, it is derived from a project given in an university OOP introductory course.
A skeleton of the project and associated unit tests were provided. 
The participants then had to design and write the game loop and logic and handle the board display to the players using the CLI.
The instructions for the participants, the skeleton project, and the projects created by the participants are available in our replication package. %

The participants used the \textsc{GitHub Copilot VScode} extension only through inline and panel completions; that is, they could not use other features of \textsc{GitHub Copilot}, such as the chat. 
They were also given access to the Internet while being forbidden from using other AI assistant (\eg ChatGPT). 

We collected \textsc{GitHub Copilot}'s telemetry by modifying the \textsc{VScode} extension and redirecting the telemetry to the computer of the participant. 
\textsc{GitHub Copilot}'s telemetry includes a plethora of different messages that enable us to retrace the history of a generation, from the moment \textsc{GitHub Copilot} decides to send a generation request to the moment of its acceptance from the user. 

The participants all used the same laptop: A Dell Latitude\,7410, with 32\,GiB of memory, an Intel Core\,i7-10610U and a CML\,GT2\,Mesa\,Intel graphics card. 
It was running on Debian (bookworm). 
The monitor was a Dell\,U2720Q.

\subsection{Dataset description}
The \textsc{AssistantTraces} dataset consists of all the telemetry sent by \textsc{GitHub Copilot} during the experiment in JSON format. 
There are $119,774$ telemetry messages in total, including $9,633$ generation requests. 
The dataset is available at \url{https://doi.org/10.5281/zenodo.11503612}.

\section{Experimental setup}\label{sec:simulation}
To estimate the energy consumption of \textsc{GitHub Copilot}, we leveraged the collected traces to simulate the developers' behavior and the front end of the code assistant on an inference server.
Specifically, the inference servers were run on a cluster, whose nodes consist of AMD EPYC\,7513 (Zen\,3), with 512\,GiB of memory and 4 Nvidia\,A100-SXM4-40GB (40\,GiB).
The server's distribution and OS were Debian 6 on Linux 5.10.0-28-amd64.

All the artifacts of this study, including our results, code, and datasets, are available in the following public repository: \url{https://doi.org/10.5281/zenodo.13167546}.

\subsection{Simulation client}
\label{subsec:client}
Thanks to the telemetry data from the \textsc{AssistantTraces} dataset, we could reproduce the API requests that \textsc{GitHub Copilot} sent to its inference server. 
We developed a simulator (the \emph{client}) to mimic developers' interactions with \textsc{GitHub Copilot} by replaying generation requests to the inference server.
The main benefit of our approach is that we can make multiple simulations with different scenarios by varying \eg the number of developers or the behavior of the code assistant.

When performing simulations with concurrent developers, we limited the simulation to the first hour of development to account for some developers who needed more than one hour to complete the task and to keep the simulation times short and manageable. 
For other developers who completed the task in less than an hour, we delayed their start times so that the middle of each simulation aligned, thus creating a more realistic distribution of the server's workload. 
For example, a 1-hour telemetry session starts at minute $0$ and reaches its midpoint after $30$ minutes, while a 50-minute session begins at minute $5$ and also reaches its midpoint at minute $30$. %

When simulating less than $20$ developers, we repeated the simulation multiple times with different developers until all were included (\eg we repeated a simulation of $2$ developers $10$ times with varying pairs of developers every time, so that all $20$ developers would be simulated). 
For simulations with more than $20$ developers, we randomly picked replicates among the $20$ developers.
To avoid issues with simultaneous identical requests, we offset each duplicate by a random number of $0$ to $30$ seconds.

\subsection{Inference server}
To handle generation requests and generate code suggestions, we set up an inference server.
Code suggestions are generated using an \glspl{llm}. 
In particular, we used the \gls{tgi} server,\footnote{\url{https://github.com/huggingface/text-generation-inference}} a server for text generation inference that is easily configurable to operate with different \glspl{llm} and that supports sharding between multiple GPUs and multiple parallel requests (using continuous batching). %
The server was set up with its default parameters, except for the number of shards, quantization method, and the number of concurrent requests which we describe in the next section.

\subsection{Studied factors}
Using the aforementioned setup, we performed several simulations with varying elements in the setup's configuration. We chose to study specific factors because we hypothesized they would affect the energy consumption of the code assistant.
The factors we studied are as follows:

    \textbf{Number of concurrent developers:} 
    We varied the number of developers concurrently querying the code assistant ranging from 1 to 500 with discrete values: 1, 2, 5, 10, 20, 30, 50, 75, 100, 150, 200, 300, 400, 500.
    
    \textbf{Streaming the requests:} 
    We emulated different request-sending behaviors from \textsc{GitHub Copilot} by activating and deactivating streaming when sending requests. 
    By default, \textsc{GitHub Copilot} uses streaming, which allows it to cancel a previous request that is still generating when a new one is sent. 
    Deactivating streaming signifies that every request that is sent by \textsc{GitHub Copilot} has to complete the triggered generation, even if it is no longer useful for the user.
    
    \textbf{Manually triggering the code assistant:} 
    Typically, \textsc{GitHub Copilot}'s generation mechanism is triggered automatically. 
    We also considered emulating a manual trigger for the generation behavior by only sending generation requests that were completed and displayed to the user based on the assumption that the developer manually triggering \textsc{GitHub Copilot} would wait for a response. 
    While this method is not perfect, it serves as an approximation of the user behavior.

    \textbf{Code assistant model:} 
    We tested three different \glspl{llm} from the StarCoder family, namely: StarCoder (15.5B parameters)~\cite{li2023starcoder}, StarCoder2-7b and StarCoder2-15b~\cite{lozhkovStarCoderStackV22024}. 
    We chose these models as they are popular open-source models for code generation. 
    The range of models allows us to see the impact of the size of the model, as well as its architecture. 
    Indeed, StarCoder is a decoder-only transformer with \emph{Multi-Query-Attention} and positional embeddings, whereas StarCoder2 is a decoder-only transformer with \emph{Grouped-Query-Attention} and \emph{Rotary-Positional-Encodings}.
    
    \textbf{Quantization method:} Quantization is a method used to reduce the computational and memory costs of running inferences.
    We studied multiple quantization methods for running the models on the inference server: EETQ~\cite{netease-fuxiNetEaseFuXiEETQ2024}, BitsAndBytes-NF4, BitsAndBytes-FP4~\cite{dettmersTimDettmersBitsandbytes2024} and no quantization at all.
    
    \textbf{Maximum number of concurrent requests:} 
    We varied the number of concurrent requests on the \gls{tgi} server which effectively modified the number of requests that could be queued simultaneously. 
    When the server was overloaded, it returned an error to the caller.
    
    \textbf{Number of GPUs:} 
    All of our servers had 4 GPUs available which enabled us to run the simulations using a subset of the available GPUs, such as two or just one GPU. 
    When using less than 4 GPUs, we considered the unused GPUs nonexistent and did not account for their consumption.

\begin{figure*}
    \centerline{\includegraphics[width=1\linewidth]{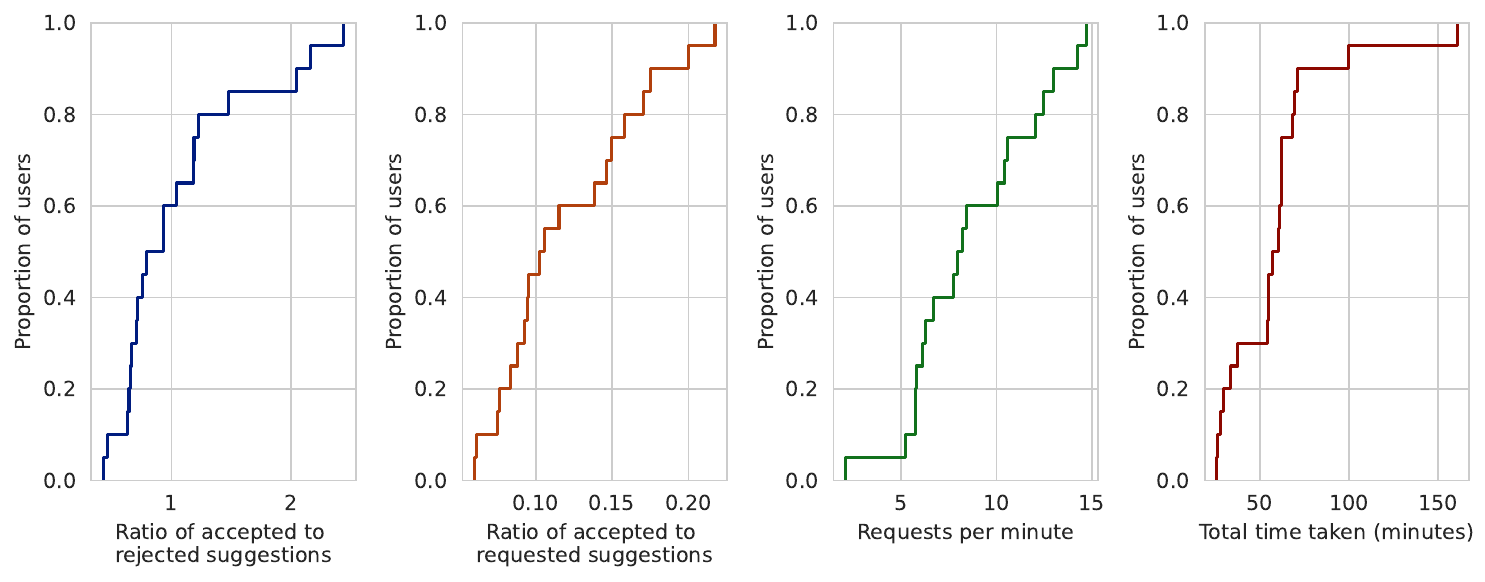}}
    \vspace{-2mm}
    \caption{Various statistics on the participants' usage of the code assistant and their time taken to finish the experiment. The first figure represents the ratio of the number of suggestions accepted by the user to the number of suggestions by the user. The second figure represents the ratio of the number of accepted suggestions to the total number of suggestions requested by the code assistant. The third one focuses on the frequency of requests made by the participant, and the last figure represents the time taken by the participant to finish the experiment.}
    \vspace{-2mm}
    \label{fig:copilot_usage_metrics}
\end{figure*}

\subsection{Configuration space exploration}
We define a configuration as a set of factors with a specific value. One example of configuration is [20 concurrent developers; streamed requests; requests are automatically triggered; the model used is StarCoder2-7b; no quantization method; 4 GPUs].

The combination of these multiple factors and their varying modalities results in 
$4,896$ unique possible configurations. 
Considering each configurations takes between 1 and 20 hours to simulate and measure, we decided not to explore the whole configuration space. 
To keep the exploration manageable yet informative, some factors were fixed while exploring the impact of others (\eg fixing the model and number of GPUs while varying the number of developers or the quantization method). 
In total, we performed $829$ simulations with $314$ unique configurations. 
As explained in \autoref{subsec:client}, configurations with less than 20 concurrent developers needed to be simulated multiple times (2 for 10 developers, 4 for 5 developers, 10 for 2 developers, 20 for 1 developer), which partly explains why there are more simulations than configurations. 
Moreover, we simulated some other configurations up to five times to measure the stability of our setup.

\subsection{Energy consumption measurements}
We measured the energy consumption of the inference server's CPU and GPU using  \texttt{perf} and \texttt{nvidia-smi} utilities, respectively.
We also collected the time taken for generations to complete (latency), the number of rejected requests (due to server saturation), and the number of completed generation requests. 
When measuring a configuration multiple times, we found that the standard deviation of the power consumption was
only $1.4$\% of the mean power consumption.
We considered this level of standard deviation acceptable and concluded that our measuring setup was stable.
To estimate the carbon emissions related to the energy consumption measured during our experiments, we considered France's 2023 carbon intensity, equivalent to 56\,g of \ensuremath{\mathrm{CO_2}} per kWh~\cite{Ember2024}.

\section{Data analysis}
\label{sec:data_analysis}

In this section, we describe our methodology for analyzing the results we obtained from \autoref{sec:simulation}.

\subsection{Simulations analysis}
\label{subsec:simulation_analysis}
To get an overview of the data, we computed the mean power consumption (in Watts) and total energy used (in Wh) for every simulation, as well as the mean latency, the number of rejected requests, and the number of completed generations. 
We also derived the power consumption per developer, which allocates an equal share of the server's energy consumption to every developer using the code assistant. 
When analyzing a simulation with multiple concurrent developers, only the period in the simulation where all developers were active in parallel was considered.

\textbf{Server saturation.}
When the server receives more requests than it can handle, the number of rejected requests or the latency may increase, depending on the client and server configuration. 
We consider that a server is saturated when the number of rejected requests exceeds 10\% or when the mean latency of the requests exceeds $20$ seconds. 
Note that this saturation threshold was set arbitrarily; searching for 
an ideal saturation threshold is out of the scope of this paper.

\textbf{Measuring the impact of a factor.}
To assess the impact of a factor on energy consumption or latency, we compared pairs of configurations that differed only by the factor being studied.
This method ensures that any observed differences are solely due to the factor in question.

\begin{figure*}
    \centerline{\includegraphics[width=1\linewidth]{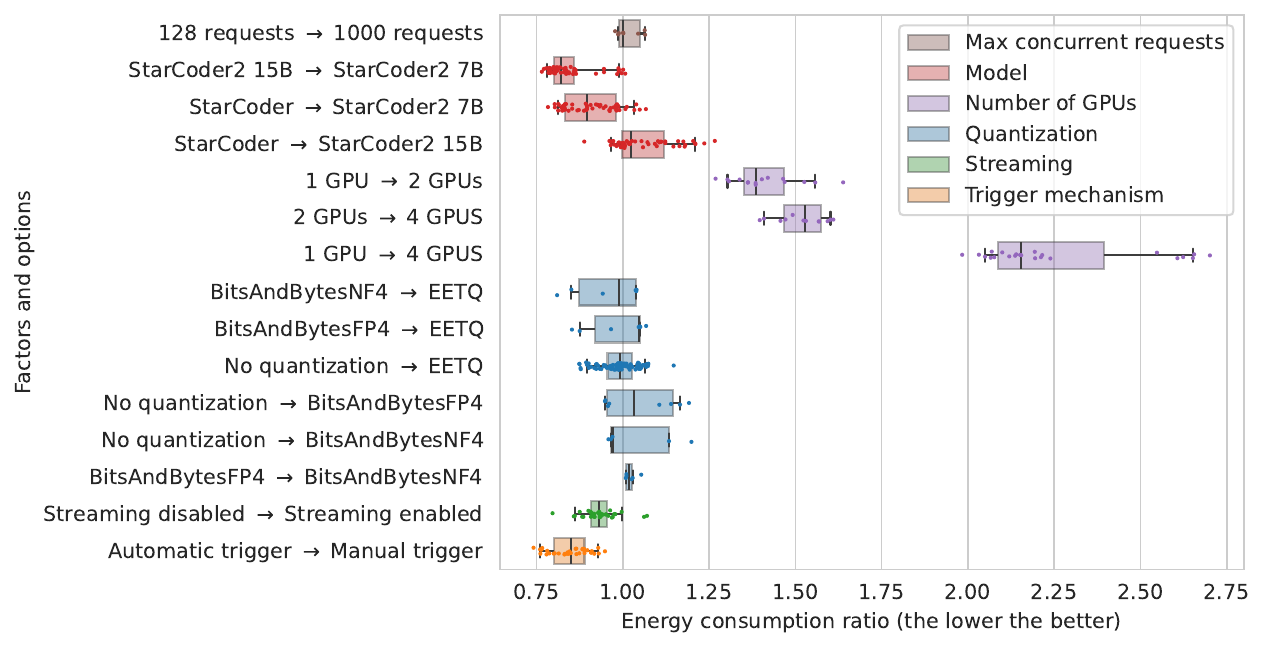}}
    \vspace{-5mm}
    \caption{Energy impact ratio from switching from one option to another. A ratio of 1 means no change, a ratio of 2 means the energy consumption doubled, and so on. Points correspond to the ratio in energy when comparing neighboring configurations.}
    \vspace{-2mm}
    \label{fig:powerdraw_impact_factors}
\end{figure*}

\begin{figure*}
    \centerline{\includegraphics[width=1\linewidth]{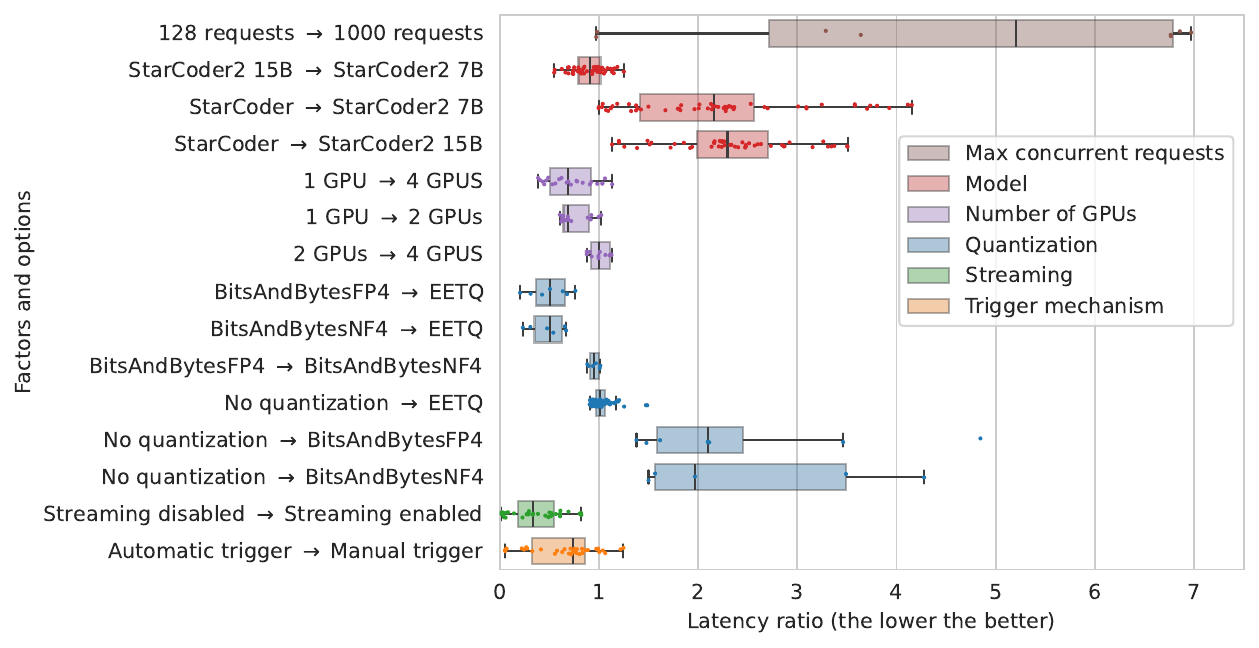}}
    \vspace{-5mm}
    \caption{Latency impact ratio from switching from one option to another. A ratio of 1 means no change, a ratio of 2 means the latency doubled, and so on. Points correspond to the ratio in latency when comparing neighboring configurations.}
    \vspace{-2mm}
    \label{fig:latency_impact_factors}
\end{figure*}

\subsection{Participant analysis}

For each participant, we performed analyses using the gathered telemetry and survey data. 
Specifically, we calculated the number of generations requested, displayed, and accepted, and those remaining in the code after a certain period. 
To determine which generation requests remain in the participant's code, we leveraged \textsc{GitHub Copilot}'s telemetry, which indicates whether a generation is still present in the code. 
This is assessed using edit distance at the character and word levels, with the generation considered removed when the word edit distance falls below 50\%~\cite{Copilotexplorera}.

In \autoref{fig:copilot_usage_metrics}, we share various statistics on the participants code assistant usage and time taken to realize the experiment. 
We observe variations in the way the participants use \textsc{GitHub Copilot}, notably in the ratio of accepted suggestions over the number of rejected suggestions or total sent generations requests. 
We also notice that some participants trigger generations more frequently than others. 
Lastly, there is also a great variation when it comes to the time the participant took to finish the experiment, indeed, $5$ participants requested to have more than one hour to finish, and $6$ completed the task within $40$ minutes.
Out of the $20$ participants, $9$ did not complete the task they were given and decided to end the experiment at the one hour mark. When calibrating the duration and complexity of the task given to the participants, we aimed at having the most development traces as possible. Thus, we chose to reduce the chances of the participants finishing early, thereby increasing the chances of the participant finishing late or giving up after one hour. We find that participants not completing the whole task is not an issue for our experiment, as the usage of the code assistant roughly stays the same during the whole task.

\section{Results}
\label{sec:results}
In this section, we summarize the key observations from our experiment and answer our research questions. 
We make our complete results available in the companion notebook in the replication package.

\begin{table*}[ht] \centering
    \begin{adjustbox}{width=1.2\textwidth,center=\textwidth}
        \begin{tabular}{lll|lrrc}
Metric                                                                                                            & Independent variable        & Groups                                & Test used & Stat   & P-value & Significant  \\ 
\midrule
\multirow{4}{*}{\begin{tabular}[c]{@{}l@{}}Ratio of accepted suggestion \\over rejected suggestions\end{tabular}} & Computer experience         & professional, student                 & t-test    & -2.496 & 0.022   & \cmark       \\
                                                                                                                  & GitHub Copilot familiarity  & regularly, sometimes, never used      & ANOVA     & 1.743  & 0.205   & \xmark       \\
                                                                                                                  & Coded Java in the last year & did code in java, didn't code in java & t-test    & -0.570 & 0.576   & \xmark       \\
                                                                                                                  & Finished                    & finished, didn't finish               & t-test    & 0.127  & 0.900   & \xmark       \\ 
\hline
\multirow{4}{*}{Requests per minute}                                                                              & Computer experience         & professional, student                 & t-test    & 0.808  & 0.429   & \xmark       \\
                                                                                                                  & GitHub Copilot familiarity  & regularly, sometimes, never used      & ANOVA     & 1.034  & 0.377   & \xmark       \\
                                                                                                                  & Coded Java in the last year & did code in java, didn't code in java & t-test    & 0.243  & 0.811   & \xmark       \\
                                                                                                                  & Finished                    & finished, didn't finish               & t-test    & 0.799  & 0.434   & \xmark      
\end{tabular}

    \end{adjustbox}
    \vspace{2mm}
    \caption{Inferential statistics of the effect of three independent variables on two metrics. The significance was asserted using a Benjamini-Hochberg correction with a False Discovery Rate of 0.20}
    \label{table:stats}
\end{table*}

\subsection{Relationship between participant's characteristics and their usage of GitHub Copilot}
Before answering any of the research questions, we wanted to have a look at the relationship between some of the participant's characteristics, such as their experience, familiarity with GitHub Copilot, if they coded in java during the last year, or whether or not they finished the task, and their usage of GitHub Copilot, that is, the rate at which the participants made requests, and the ratio of code suggestions they accepted when presented to them. We performed this investigation in order to find if any noticeable effect could bias our subsequent evaluations, so the metrics were chosen because they become important later on. In total, we performed 8 post-hoc tests, which are described in \autoref{table:stats}. In order to reduce the Type I error rate (false positives), we also performed a Benjamini-Hochberg procedure to determine the significance of the tests using a false-discovery rate (FDR) of 0.20.

From the data in \autoref{table:stats}, we cannot conclude that there is any effect between most of the characteristics studied and the studied metrics, except between the computer experience and the ratio of accepted suggestions over rejected suggestions. When analyzing the data, we see that professional developers have a ratio of 0.67 whereas students have a ratio of 1.26. This indicates that the professional developers in our sample are more conservative towards the suggestions made by GitHub Copilot compared to the students. The students tend to accept the suggestions made by Copilot twice as much as the professional developers.

\subsection{RQ1: What is the impact of studied factors on the energy consumption and performance of code assistants ?}\label{sec:rq1}
In this section, we report on our findings on the different studied factors and their impacts. 
In \autoref{fig:powerdraw_impact_factors} and \autoref{fig:latency_impact_factors}, we depict the impact of switching from one option to another on energy consumption and request latency, respectively. 
The impacts were calculated using the method described in \autoref{subsec:simulation_analysis}. 
Each hue represents one factor, while each row represents the switch from one option to the next. 
The X-axis reflects the ratio in measured latency or energy consumption between the first and second options.

\begin{figure}%
    \centerline{\includegraphics[width=\linewidth]{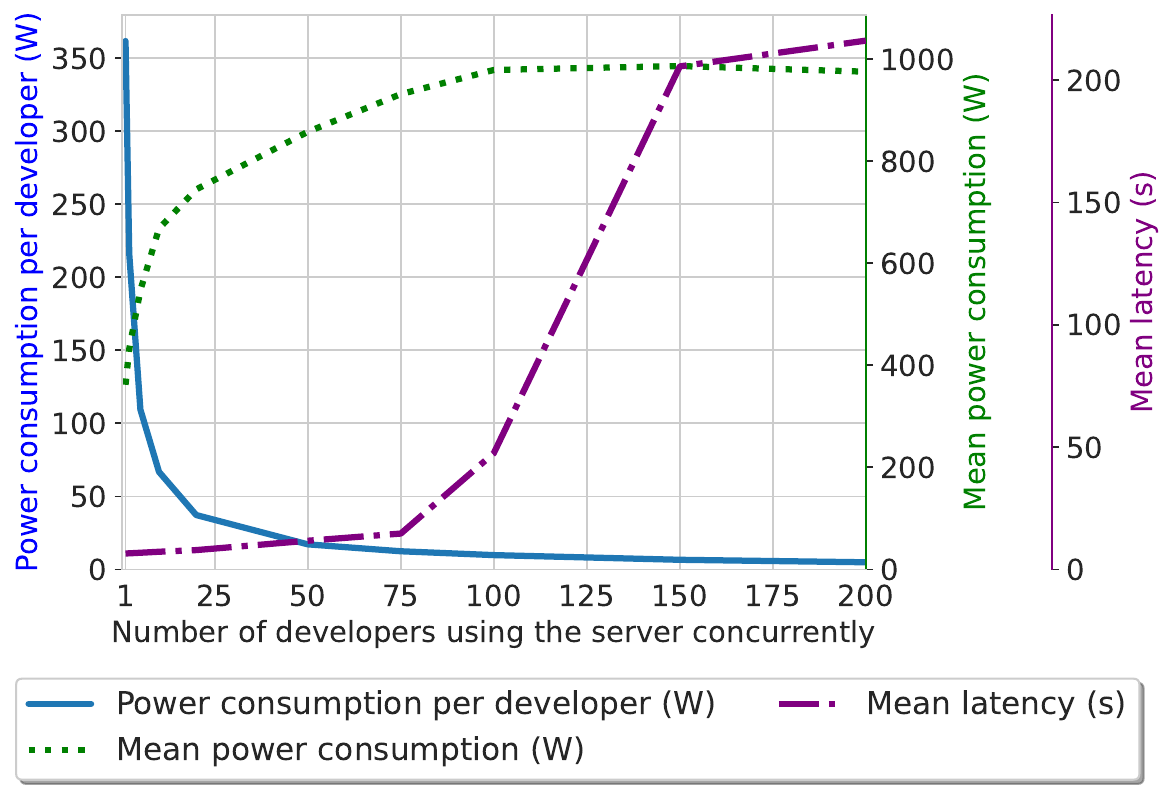}}
    \caption{Evolution of the average power consumption and the latency depending on the number of developers, for the following configuration: StarCoder2-7B; no quantization; no streaming; manual trigger; maximum 1000 concurrent requests; 4 GPUs. The latency and power consumption are superposed in order to easily visualize how they interact when the number of developer increases.}
    \label{fig:powerdraw_per_dev}
\end{figure}

\textbf{Number of concurrent developers.}
When increasing the number of concurrent developers, we observe an increase in the average power consumption of the machine and the latency of the server up to a certain point. 
Thanks to the continuous batching techniques used by \gls{tgi}, adding more developers only marginally increases the latency and consumption of the server, thus reducing the energy consumption per developer. 
This is illustrated in \autoref{fig:powerdraw_per_dev}, where we can observe that the energy consumption per developer decreases whenever a developer is added. 
On the other hand, with too many developers, the latency becomes excessively high, making the code assistant unusable. 
With the configuration shown in \autoref{fig:powerdraw_per_dev}, the average latency reaches $16$ seconds with $75$ concurrent developers, and increases exponentially from that point with each developer added ($50$ seconds at $100$ developers, $210$ seconds at $150$ developers, \etc). 
It finally reaches a plateau past $150$ concurrent developers as the server reaches the maximum number of concurrent requests limit and starts rejecting new requests.

\textbf{Streaming the requests.}
Enabling streaming to send requests---i.e., canceling previous generation requests when a new one arrives---reduces server latency by $62$\%, on average, and reduces power consumption by $7$\%.
The lesser reduction in power consumption compared to the reduction in latency is due to the inference server still spending most of its time generating responses.
As a result of the lower latency, streaming allows more concurrent developers to use the assistant.

\textbf{Manually triggering the code assistant.}
Manually triggering the code assistant by only requesting the generations that were proposed to the developer, reduces energy consumption by $15$\% and reduces latency by $35$\%. 
The highest reduction in energy consumption ($25$\%) is observed with the configuration [StarCoder2; no quantization; no streaming automatic trigger; maximum 1000 concurrent requests; 2 GPUs; 5 developers], which also reduces the latency by $25$\%. On the other hand, the lowest reduction of the energy consumption ($5$\%) is observed with the configuration [StarCoder2-7b; no quantization; no streaming; automatic trigger; maximum 1000 concurrent requests; 4 GPUs; 75 developers] which, however, reduces the latency by $93$\%. 
In this specific configuration, enabling an automatic trigger made the server saturate ($226$ seconds of latency and $59$\% of rejected requests), while using a manual trigger allowed it to handle the load from the $75$ developers ($14.6$ seconds of latency and $0$\% of rejected requests).  

\textbf{Quantization.}
Quantization generally increases latency.
Notably, employing the method {\sf EETQ} over no quantization at all increases the latency by $3.1$\%, and using {\sf BitsAndBytes-NF4} or {\sf BitsAndBytes-FP4} increases the latency by $138.6$\% and $156$\%, respectively. 
In some cases, the use of {\sf EETQ} also reduces energy consumption: on average, it reduces energy consumption by $1.1$\%, but it can reduce it by up to 12.6\% in the case of the configuration [StarCoder; streaming; automatic trigger; maximum 1000 concurrent requests; 4 GPUs; 5 developers]. 
{\sf BitsAndBytes}, however, slightly increases the power consumption by 5\% on average.

\textbf{Model.}
Using \glspl{llm} with fewer parameters reduces energy consumption and latency.
For instance, switching from StarCoder2-15B to StarCoder2-7B reduces energy consumption by $15.6$\% and latency by $10.0$\%.
However, the model architecture also plays an important role in latency. 
For example, while StarCoder (15.5B) and StarCoder2-15B exhibit a similar number of parameters and power consumption, using the latter over the former increases the latency by $149$\% on average. 
We believe this might be due to differences in the architecture of the two models.
For instance, StarCoder uses {\em Multi-Query-Attention}, whereas StarCoder2 adopts {\em Grouped-Query-Attention}.

\textbf{Maximum number of concurrent requests.}
Increasing the maximum number of concurrent requests allowed by the \gls{tgi} server does not affect the energy consumption of the server, but greatly increases latency when there are too many concurrent developers---\ie up to 597\% increase with the configuration [StarCoder2-7B; EETQ; no streaming; automatic trigger; 4 GPUs; 50 developers]. 
Providers should prefer having a lower maximum to reject the requests early rather than making the users wait for several minutes.

\textbf{Number of GPUs.}
As intuitively expected, reducing the number of GPUs allocated to the inference server reduces the energy consumption of the server and increases latency. 
For example, using 4 GPUs instead of 2 increases consumption by 51.8\% and can decrease the latency in some cases, hence enabling some more concurrent developers to use the code assistant as more memory is available. 
Downsizing the number of GPUs can be a viable option for saving energy if the number of concurrent developers is low enough.

\textbf{Usage of the code assistant.}
Participants' usage of \textsc{GitHub Copilot} varied, with the average amount of requests per minute ranging from $1.9$ to $14.7$, overall averaging $9$ requests per minute. 
We observe that the more a developer uses \textsc{GitHub Copilot} (\ie, the more generation requests are made), the more power consumption increases (correlation of $0.93$). 
We infer that in its current state, the usage of a code assistant is directly correlated with the amount of code a participant writes---\ie the more time a participant spends writing code in the \gls{ide}, the more a code assistant triggers generations.

\begin{samepage}
\begin{formal}
\textbf{RQ1}: Various factors significantly impact the energy consumption and performance of code assistants. 
Increasing the number of concurrent developers improves energy efficiency, but risks server saturation. 
Streaming requests and manually triggering generations both significantly reduce energy consumption and latency. 
Quantization methods and the choice of \gls{llm} also affect performance, with smaller models showing better efficiency. 
Adjusting the number of GPUs is crucial for optimizing energy use.
Our findings, therefore, confirm the importance of carefully selecting and optimizing these factors.
\end{formal}
\end{samepage}

\subsection{RQ2: How many generation requests made by \textsc{GitHub Copilot} are actually useful?}

\begin{figure}[t!]
    \centerline{\includegraphics[width=\linewidth]{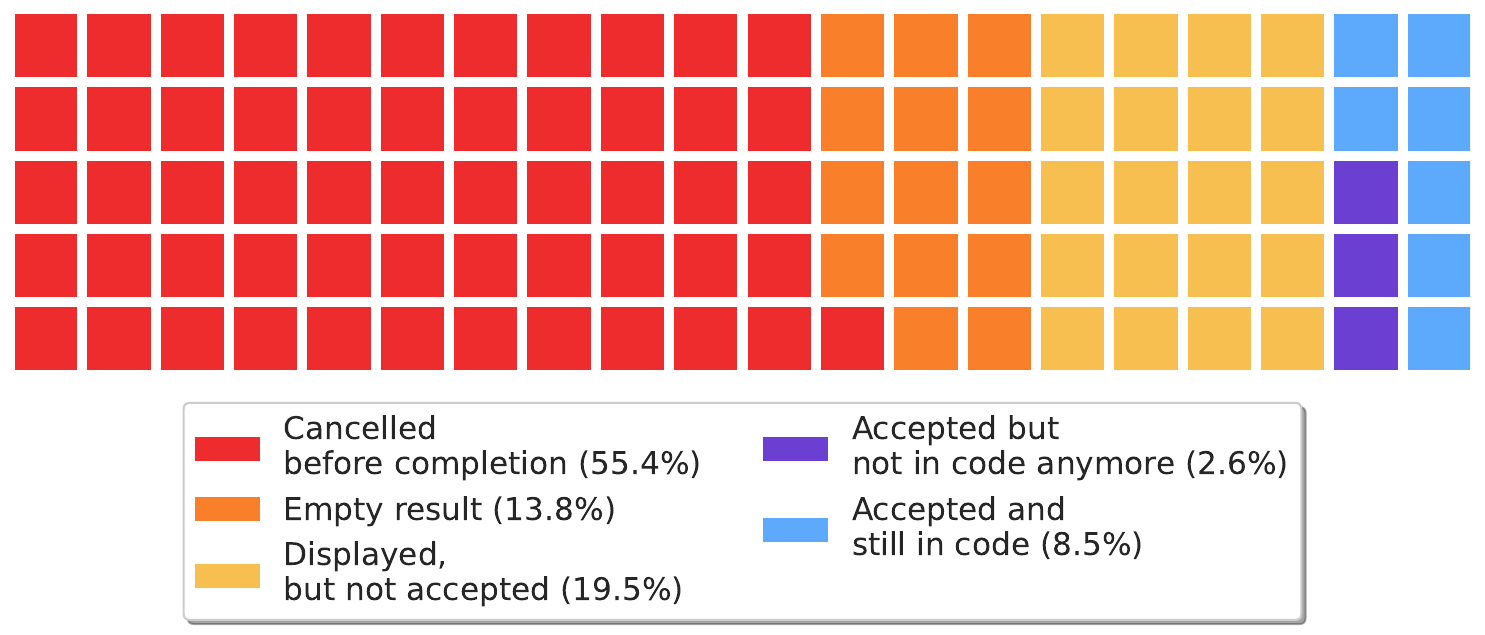}}
    \caption{Percentage of requests sent by \textsc{GitHub Copilot} depending on their completion state. Red-tinted categories represent requests that did not benefit the user. Blue-tinted categories represent requests that benefited the user.}
    \label{fig:requests_proportion}
\end{figure}

\autoref{fig:requests_proportion} illustrates the final state of generation requests sent by \textsc{GitHub Copilot} during our experiment.
Out of the $9,634$ generation requests made by \textsc{GitHub Copilot} over the $20$ participants, only $2,944$ finished generating and were displayed to the user (30\%). 
Of these, $1,066$ were accepted (11\% of all requests), and $816$ (8.5\%) were kept in the code at the end of the experiment according to \textsc{GitHub Copilot}. 

Our analysis reveals that the majority of generation requests made by \textsc{GitHub Copilot} were canceled, often because they were started while the user was still typing, leading to new requests that replaced the previous ones. 
Empty completions typically occur at the end of a sentence, or before a closing parentheses or brackets. 
The completions that were displayed but not accepted are most likely due to either the users not wanting or needing the suggestion in the first place, or to the suggestions not matching what the user expected.

Previously, we discussed the energy implications of these inefficiencies when presenting the energy saving of manually triggering \textsc{GitHub Copilot}. 
By eliminating unnecessary requests, \ie canceled and empty requests, we could save 15\% of energy on average. 
This relatively modest energy saving, despite removing almost 70\% of the generations is due to the lower-than-average computing time used by canceled and empty generations, compared to completed generations. 
Assessing the energy impact of suggestions displayed, but not accepted, by users is left for future research.

\begin{samepage}
\begin{formal}
\textbf{RQ2}: An analysis of \textsc{GitHub Copilot}'s generation requests reveals a significant share of them are not useful to the end user, indicating that many generation requests are either unnecessary or not helpful to users. 
This inefficiency highlights the potential for substantial energy savings by optimizing when and how generation requests are made, possibly through more intelligent triggering mechanisms or better user interaction designs.
\end{formal}
\end{samepage}

\subsection{RQ3: Under different scenarios and objectives, how much does a developer using a code assistant such as \textsc{GitHub Copilot} cost in energy?}
\label{subsec:res_req3}

For the sake of simplicity, this section only focuses on a subset of configurations and their impacts, given the large number of different configurations available ($314$).
However, the entire set of configurations can be found in our companion notebook.
When presenting these results, we assume that the server load is constant---\ie all developers are working simultaneously---and that the server is turned on only when the developers are working.

To select the configurations, we defined three different development scenarios, each with two objectives. 
The goal of these scenarios is to highlight how different use cases require different configurations and have varying impacts. The scenarios are as follows:
\begin{itemize}
    \item \textbf{Small team:} 5 developers concurrently requesting a single dedicated server machine.
    \item \textbf{Medium team:} 20 developers concurrently requesting a single dedicated server machine.
    \item \textbf{Distributed service (\eg \textsc{GitHub Copilot}):} A large number of developers (sharing) requesting many server machines. The aim is to maximize the number of developers per machine while not saturating the servers. We report on the impact of a single machine in this scenario. 
\end{itemize}

The two objectives a development team may aim for we considered in our study are as follows:
\begin{itemize}
    \item \textbf{Performance:}  By design, the generation requests are triggered automatically, and the previous ones are canceled (using streaming). The latency between requests is thus minimized. Generations are triggered automatically so that the developer always receives suggestions as quickly as possible. This is also the default triggering and streaming mode of \textsc{GitHub Copilot}.
    \item \textbf{Frugality:} The energy consumption per developer is minimized. The generation requests are triggered manually, and the previous ones are not canceled (not using streaming), as we expect the developers to wait for their manually triggered generation requests to finish.
\end{itemize}

\begin{figure}[t!]
    \centerline{\includegraphics[width=\linewidth]{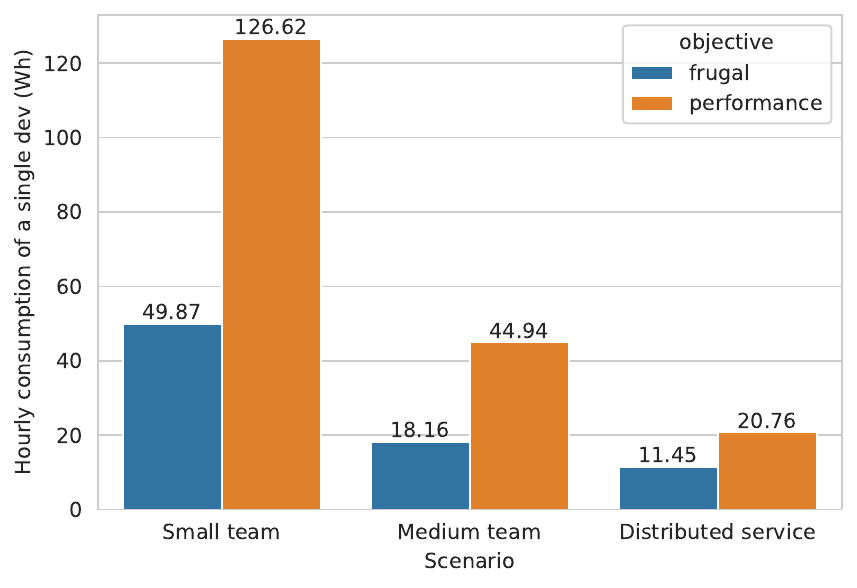}}
    \vspace{-4mm}
    \caption{Hourly energy consumption (Wh) of a single developer based on the objective and the number of developers using the assistant.}
    \label{fig:energy_per_hour}
\end{figure}

\autoref{table:best_config} shows the best configurations for each scenario and objective. 
For each of the $6$ configurations, its latency and various energy consumption metrics are presented. 
To better put in perspective the energy usages of the configurations, we also show them in \autoref{fig:energy_per_hour}.
As we can see from both the table and the figure, the energy consumption of a single developer varies significantly across the different configurations. 
The setup machine's high idle power consumption (\textasciitilde270\,W for four GPUs and one CPU) results in a substantial per-developer energy impact when fewer developers utilize many GPUs. 
Conversely, increasing the number of concurrent developers reduces individual energy consumption by leveraging the continuous batching from the \gls{tgi} server, aligning with the findings discussed in \citesec{rq1}.

If \textsc{GitHub Copilot} were using a setup similar to ours with maximized server load, the average power consumption per developer could range from 10\,W to 20\,W depending on their objective.
However, with dedicated servers for small teams, under-utilization could result in a high power consumption per developer (\textasciitilde120\,W). 
This can be mitigated by using fewer GPUs and smaller models, or considering server sharing.

To put it in perspective, we measured the energy of the laptop and the monitor of the last two participants using a power-meter. 
The average power consumption of the laptop and monitor when used by our participants was 19.7\,W and 18\,W, respectively. 
Thus, in \textsc{GitHub Copilot}'s best-case scenario using our setup, the power consumption per developer would be about 58\% the consumption of a laptop's consumption, and at worst it would be roughly equal to that of a laptop.

We observe that all performance configurations uses StarCoder due to its low latency. 
However, among the three studied models, StarCoder is the worst at generating code: StarCoder has a 33.6 pass@1 on HumanEval while StarCoder2-7B and 15B have 35.4 and 46.4 pass@1~\cite{li2023starcoder, lozhkovStarCoderStackV22024}, respectively.

\begin{table*}[ht] \centering
    \renewcommand{\arraystretch}{1.2} %
    
    \begin{adjustbox}{width=1.3\textwidth,center=\textwidth}
        \begin{tabular}{l|cc|cc|cc}
\toprule
\bf Scenario & \multicolumn{2}{c}{\makecell{{\bf Small team} \\({\em dedicated server})}} & \multicolumn{2}{c}{\makecell{{\bf Medium team} \\({\em dedicated server})}} & \multicolumn{2}{c}{{\bf Distributed service}} \\
Objective & frugal & performance & frugal & performance & frugal & performance \\
\midrule
Number of concurrent developers             & 5 & 5 & 20 & 20 & 75 & 50 \\
Model                                       & StarCoder2-7B & StarCoder & StarCoder2-7B & StarCoder & StarCoder2-7B & StarCoder \\
Quantization method                         & - & - & - & - & EETQ & - \\
Number of GPUs                              & 1 & 4 & 1 & 4 & 4 & 4 \\
Streaming                                   & \xmark & \cmark & \xmark & \cmark & \xmark & \cmark \\
Manual trigger emulation                    & \cmark & \xmark & \cmark & \xmark & \cmark & \xmark \\
\cline{1-7} Average latency (s)             & 6.4 & 1.2 & 7.7 & 1.7 & 16.5 & 2.3 \\
Average server power (W)                    & 249.3 & 633.1 & 363.2 & 898.8 & 858.9 & 1038.2 \\
Energy per 1000 generation requests (Wh)     & 12.0 & 30.7 & 0.9 & 2.2 & 0.1 & 0.4 \\
Energy per hour per developer (Wh)          & 49.9 & 126.6 & 18.2 & 44.9 & 11.4 & 20.8 \\
CO2 emissions per hour per developer (g)    & 2.8 & 7.1 & 0.9 & 1.0 & 0.6 & 1.2 \\
\bottomrule
\end{tabular}

    \end{adjustbox}
    \vspace{2mm}
    \caption{Optimal configurations for each scenario and objective, and their energy and performance impacts.}
    \vspace{-5mm}
    \label{table:best_config}
\end{table*}

\begin{samepage}
\begin{formal}
\textbf{RQ3}: The energy cost for a developer using a code assistant like \textsc{GitHub Copilot} varies significantly based on its configuration. Small teams with dedicated servers can see high per-developer energy consumption (\textasciitilde 120\,W), while large-scale services can achieve much lower consumption (10--20\,W) by maximizing server load. 
Overall, our results emphasize the importance of choosing appropriate configurations based on team size and objectives to optimize both energy consumption and performance when using code assistants.
\end{formal}
\end{samepage}

\section{Discussion and implications}
\label{sec:discussion}
In this section, we discuss the results reported in the previous section and their implications.

\textbf{On energy usage.}
As reported in our results, the configuration choice for a code assistant significantly impacts per-developer power consumption.
However, a considerable portion of the energy spent on generations is wasted due to cancellations and disinterest from users. 
Substantial energy savings could be achieved by requiring users to manually trigger \textsc{GitHub Copilot} and encouraging them to rethink their interactions with the tool, or by enhancing the request-triggering mechanism. 
For instance, Mozannar~\etal~\cite{mozannarWhenShowSuggestion2024} proposed a novel method for triggering generation requests by leveraging human feedback collected from \textsc{GitHub Copilot}'s telemetry. 
Furthermore, Barke~\etal~\cite{barkeGroundedCopilotHow2022} showed that developers exhibited two kinds of behaviors when interacting with \textsc{GitHub Copilot}---\textit{acceleration mode} and \textit{exploration mode}---which could also be leveraged to adapt the triggering mechanism of \textsc{GitHub Copilot}.

One of our objectives with this paper is to raise awareness among practitioners, tool providers and researchers about the environmental impact of using code assistants. 
Developers using code assistants can significantly reduce their energy usage by manually triggering the generations and using the assistant in a more conscious manner to avoid unnecessary generations. 
On the other hand, the providers of the assistants should maximize the number of users per inference server and carefully select the quantization method so as to decrease the resource usage. The choice of the model is also paramount, as it dictates the amount of memory left for batching requests, and the computational cost of a single request.

We encourage researchers to pursue our work by finding more ways to enable users and providers to reduce the energy consumption of their code assistants. 
One avenue of work is to evaluate the environmental footprint of the assistant using life-cycle analysis, in order to incorporate the hardware and training costs. 
Lastly, while code assistants and more specifically GitHub Copilot can improve developer productivity~\cite{MeasuringGitHubCopilot2024}, the rebound effects~\cite{dimitropoulosEnergyProductivityImprovements2007} stemming from faster development needs to be assessed. 
We hope that our dataset, \textsc{AssistantTraces}, will serve as a valuable resource for researchers and tool providers. 
We encourage researchers to build upon \textsc{AssistantTraces}, enhance its collection methodology, and refine it through their own user studies.

\textbf{On the number of developers.}
Our results indicate that using a inference server for a small number of developers can be inefficient in terms of hardware and energy. 
Optimizing the number of developers per machine improves utilization, but a balance is needed between developer load and code assistant performance.
Having few developers complicates hardware and model selection, as smaller GPUs may struggle with large models, while smaller models may generate lower-quality code. 
In this case, practitioners should consider sharing hardware and its costs with others, so that they may all benefit from higher-end hardware and a lower individual impact. 
We also encourage researchers to develop and improve sharing models for \gls{llm} inference, such as Petals~\cite{borzunov2022petals}.

\textbf{On the difference in latency between StarCoder models.}
We observed a sharp increase in latency between the StarCoder and StarCoder2-15B models. 
While we believe this is due to the architecture changes introduced by the StarCoder2 model~\cite{lozhkovStarCoderStackV22024}, more research is needed to get a better insight into the reasons for such an increase. 
As latency plays a key role in the maximum number of concurrent developers an inference server can handle, finding ways to reduce said latency could help further improve the efficiency of code assistants.

\textbf{On quantization.} 
In our results, we found that quantization effectively increased the latency and had close to no effect on the energy consumption. 
We find this result surprising considering quantization reduces the \gls{llm}'s size. 
It is possible that these specific quantization method increase the computational cost of inference. 
We advise researchers to investigate the reasons for this effect, and explore the energy savings of other quantization methods.

\textbf{On the generalization of our results.}
Our study uses one specific setup and varies its configuration. While the energy consumption we observed in our different scenarios give a good idea of the scale of the energy required to run a code assistant, the observed energy consumption is specific to the context of our experiment, and should not be generalized to other code assistants, especially \ghc. However, even though other code assistants might use a different hardware, model or serving framework, they can still benefit from our results. Another setup, completely different from ours, with users making a different amount of requests per minute (which is highly correlated with the energy consumption of the assistant) would still save energy by maximizing the number of concurrent developers per machine, by using smaller models, by having the users manually trigger the generations and by cancelling them early if they're not needed anymore, \etc

\textbf{On the environmental footprint of the code assistants}
We only calculated the CO2 emissions that were due to the energy consumption of the inference server. Future studies could also include the other sources of impact such as the hardware the server is hosted on, and the datacenter its in. This could also be an opportunity to present other metrics such as resource depletion or water use, which can be as important as CO2 emissions when talking about ICT.~\cite{simonUncoveringEnvironmentalImpact2023}

\section{Limitations}\label{sec:limits}
In this section, we address several limitations of our study:

\textbf{Participants and task selection.}
Selecting participants by word of mouth, and mailing lists from university students and professionals allowed us to easily recruit enough qualified participants for our study. 
Moreover, the task assigned to the participants was designed to be short and simple for an experimental setting, yet long and complex enough to simulate a realistic development session. 
One alternative was to perform our dataset collection on developers working for their companies or school projects. 
While this alternative could have provided us with more realistic development traces, it would have been logistically harder to perform. 
Another alternative was to assign multiple short tasks, following Vaithlingam~\etal~\cite{vaithilingamExpectationVsExperience2022}, which could simplify generalization, while compromising the realism of the full development process, including including design, debugging, and testing.

Having a non-representative sample of the population of developers and a not complex-enough task may affect some of our results such as the number of accepted/rejected generation requests, as well as the number of requests per minute made by the participants. As the rate of generation requests is highly correlated (0.93) with the energy consumption of code assistant, the raw energy consumption numbers would also be affected. If our sample was indeed biased, we believe it would not affect the results from RQ1 and RQ3, as our experiment focuses mainly on the impact of changing configuration options of the code assistant. Another setup, completely different from ours, with users making a different amount of requests per minute (which is highly correlated with the energy consumption of the assistant) will still save energy by maximizing the number of concurrent developers per machine, by using smaller models, by having the users manually trigger the generations and cancel them early if they're not needed anymore, \etc

\textbf{Configuration space and exploration.}
To keep the number of simulations low, we had to limit both the configuration space's size and its exploration. 
Indeed, we restricted our evaluation to three models from the StarCoder family and ran our simulations without varying the hardware. 
Additionally, we only considered a subset of \gls{tgi} parameters. 
As a result, some configurations yielding different or optimal results might not be explored. 
We mitigated this effect by curating the configurations to explore so that they cover a large spectrum.

\textbf{Emulation of a manual trigger of a code assistant.}
Our method of emulating the manual triggering of \textsc{GitHub Copilot} by only sending generation requests displayed to the user is only an approximation that allowed us to save time and resources by simplifying our dataset collection process.
In reality, user behavior may differ, and this emulation might not fully capture the nuances of manual interactions. 
A dedicated study involving real users manually triggering generations is needed to obtain more accurate insights.

\textbf{Simulation of the developers.}
Simulating the developers instead of performing live measures allowed us to have more freedom and explore more configurations.
In return, our simulations do not allow for any kind of change in behavior from the simulated developer---\ie the same request will be sent at the same times no matter the latency or quality of the responses. 
An alternative was to measure the energy as the participants were developing and modify \textsc{GitHub Copilot} to use our inference server. 
While this would yield more accurate reactions from the developers, it would greatly limit the amount of configurations we could explore. 
A compromise was to develop autonomous agents simulating the developers, which we deemed to difficult to realize.

\textbf{Quality of the generated code.}
When performing the simulations, we decided not to assess or take into consideration the quality or correctness of the code generated by the \glspl{llm}. 
It was possible to estimate the quality of the code by comparing it to the code written to the developer using a \emph{BiLingual Evaluation Understudy} (BLEU) score~\cite{papineniBLEUMethodAutomatic2002}. 
However, as shown by Chen~\etal~\cite{chenEvaluatingLargeLanguage2021}, any score relating to the proximity between two texts is unreliable at best when evaluating \glspl{llm} for code. 
Another alternative was to evaluate the generations by hand, which would have been too much time intensive. 
A third alternative was to consider the reported pass@k from the model's creators, which could have been easy to implement, but would have made our search of optimal configuration harder in \autoref{subsec:res_req3}.
The rationale for our choice can be explained by the lack of methods to evaluate the quality of the generated code in the context of our experiment and the simplicity of not considering the code quality when searching for the best models under specific scenarios and objectives. 
Nevertheless, the trade-off between energy consumption, performance, and code quality is a consideration that should be addressed in future research.

\section{Related Works}\label{sec:rw}\label{sec:related}
Previous research have investigated various properties of \glspl{llm} specialized in code generation:
previous research has investigated various aspects of \glspl{llm} for code-related tasks, including the security of their suggestions~\cite{pearceAsleepKeyboardAssessing2022, sandovalSecurityImplicationsLarge2022, perryUsersWriteMore2022}, the prevalence of bugs in the generated code~\cite{jesseLargeLanguageModels2023}, how developers interact with them ~\cite{barkeGroundedCopilotHow2022, vaithilingamExpectationVsExperience2022, perryUsersWriteMore2022} or just the quality and correctness of the code they generate~\cite{doderleinPilotingCopilotCodex2022, yetistirenAssessingQualityGitHub2022,  nguyenEmpiricalEvaluationGitHub2022,  liu2024your}. 
There have also been efforts to measure the efficiency of \glspl{llm} through the creation of benchmarks for comparing them, such as {\sc HumanEval}~\cite{chenEvaluatingLargeLanguage2021}, MBPP~\cite{austinProgramSynthesisLarge2021},  {\sc CoderEval}~\cite{yuCoderEvalBenchmarkPragmatic2023}, APPS~\cite{hendrycksMeasuringCodingChallenge2021}, {\sc CodeXGLUE}~\cite{luCodeXGLUEMachineLearning2021} or {\sc ReCode}~\cite{wangReCodeRobustnessEvaluation2022}. 
Xu~\etal~\cite{xuSystematicEvaluationLarge2022} also conducted a comparative evaluation of multiple \glspl{llm} for code.

Specifically, researchers have also investigated the energy consumption and carbon footprint of various aspects of Deep Learning, such as computer vision~\cite{luccioniPowerHungryProcessing2023, luccioniCountingCarbonSurvey2023, selvanEquityAccessCase2024} and natural language processing~\cite{luccioniPowerHungryProcessing2023, vartziotisLearnCodeSustainably2024}.
Notably, in the context of \glspl{llm}, Vartziotis~\etal studied the \textit{Green Capacity} of \glspl{llm} for code~\cite{vartziotisLearnCodeSustainably2024}. 
They quantified the sustainability awareness of ChatGPT, \textsc{GitHub Copilot} and CodeWhisperer, based on the energy consumption and performance of the solutions they generated for Leetcode problems. 
Coignion~\etal studied the performance of the code generated by multiple \glspl{llm} on Leetcode, and found that the \glspl{llm} studied had similar code performance~\cite{coignion:hal-04525620}. 
Luccioni~\etal measured the inference cost in energy and carbon of multiple \glspl{llm}, and found that multi-purpose \glspl{llm} are orders of magnitude more expensive than specialized \glspl{llm}~\cite{luccioniPowerHungryProcessing2023, luccioniCountingCarbonSurvey2023}. 

Samsi~\etal benchmarked the inference performance and energy costs of different sizes of LLaMa models on two GPU types~\cite{samsiWordsWattsBenchmarking2023}. 
The study provided the words per second performance of the LLaMa models, as well as the energy per second, per decoded token and per response. 
Multiple factors were studied, such as the dataset used to perform the generations, the batch size, the type of GPU and the length of the generation.
Chakravarty~\etal analyzed the effect of quantization on the energy efficiency, accuracy, memory usage, and speed of speech recognition models~\cite{chakravartyDeepLearningModels2024}.

Our study stands out by specifically examining the energy consumption of \gls{llm}-based code assistants during real-world developer interactions. Unlike previous research that focused on \gls{llm} training impacts or isolated inference impacts, we analyze the energy usage of code assistants in a practical setting. Our dataset and methodology allow us to explore various factors influencing energy consumption, such as the number of concurrent developers, offering insights for service providers and end-users to optimize resource usage. By addressing these gaps and highlighting practical implications, our research contributes to Green AI, making it one of the first investigations into the energy consumption of code assistants.

\section{Conclusion}\label{sec:conclusion}
In this study, we explored the energy consumption of code assistants powered by \glspl{llm}, focusing on \textsc{GitHub Copilot}. 
Our findings indicate that various configuration factors, such as the number of concurrent developers, model size, and use of streaming, impact both the energy consumption and performance of these tools. 
Notably, a substantial amount of energy is spent on generation requests that are ultimately unused, highlighting an area for potential improvement.

Our results indicate that significant energy savings can be achieved by service providers and end-users if they take the following steps: disable the automatic triggering of code assistant suggestions on the client side, utilize batching techniques to maximize inference server usage and support more concurrent developers, and reduce model resource consumption through effective quantization and the use of more efficient models.

Overall, while \gls{llm}-based code assistants offer productivity benefits, it is crucial to consider their environmental impact. 
By adopting more efficient configurations and usage practices, we can make these tools more sustainable without compromising their utility.

\section*{Acknowledgment}
This work received support from the French government through the {\em Agence Nationale de la Recherche} (ANR) under the France\,2030 program, including partial funding from the {\sc CARECloud}~({\sf ANR-23-PECL-0003}), DISTILLER~({\sf ANR-21-CE25-0022}), and KOALA~({\sf ANR-19-CE25-0003-01}) projects.
Experiments presented in this paper were carried out using the Grid'5000 testbed, supported by a scientific interest group hosted by Inria and including CNRS, RENATER and several Universities as well as other organizations.\footnote{See \url{https://www.grid5000.fr}}

\newpage

\bibliographystyle{elsarticle-num} 
\bibliography{biblio.bib}

\begin{thebibliography}{10}
\expandafter\ifx\csname url\endcsname\relax
  \def\url#1{\texttt{#1}}\fi
\expandafter\ifx\csname urlprefix\endcsname\relax\def\urlprefix{URL }\fi
\expandafter\ifx\csname href\endcsname\relax
  \def\href#1#2{#2} \def\path#1{#1}\fi

\bibitem{schwartzGreenAIPaper2020}
R.~Schwartz, J.~Dodge, N.~A. Smith, O.~Etzioni, Green {{AI}}, Communications of
  the ACM 63~(12) (2020) 54--63.
\newblock \href {https://doi.org/10.1145/3381831} {\path{doi:10.1145/3381831}}.

\bibitem{samsiWordsWattsBenchmarking2023}
S.~Samsi, D.~Zhao, J.~McDonald, B.~Li, A.~Michaleas, M.~Jones, W.~Bergeron,
  J.~Kepner, D.~Tiwari, V.~Gadepally, From {{Words}} to {{Watts}}:
  {{Benchmarking}} the {{Energy Costs}} of {{Large Language Model Inference}},
  in: 2023 {{IEEE High Performance Extreme Computing Conference}} ({{HPEC}}),
  2023, pp. 1--9.
\newblock \href {https://doi.org/10.1109/HPEC58863.2023.10363447}
  {\path{doi:10.1109/HPEC58863.2023.10363447}}.

\bibitem{luccioniPowerHungryProcessing2023}
S.~Luccioni, Y.~Jernite, E.~Strubell, Power {{Hungry Processing}}: {{Watts
  Driving}} the {{Cost}} of {{AI Deployment}}?, in: The 2024 {{ACM Conference}}
  on {{Fairness}}, {{Accountability}}, and {{Transparency}}, ACM, Rio de
  Janeiro Brazil, 2024, pp. 85--99.
\newblock \href {https://doi.org/10.1145/3630106.3658542}
  {\path{doi:10.1145/3630106.3658542}}.

\bibitem{vartziotisLearnCodeSustainably2024}
T.~Vartziotis, I.~Dellatolas, G.~Dasoulas, M.~Schmidt, F.~Schneider,
  T.~Hoffmann, S.~Kotsopoulos, M.~Keckeisen, Learn to {{Code Sustainably}}:
  {{An Empirical Study}} on {{LLM-based Green Code Generation}} (Mar. 2024).
\newblock \href {http://arxiv.org/abs/2403.03344} {\path{arXiv:2403.03344}}.

\bibitem{coignion:hal-04525620}
T.~Coignion, C.~Quinton, R.~Rouvoy, {A Performance Study of LLM-Generated Code
  on Leetcode}, in: {EASE'24 - 28th International Conference on Evaluation and
  Assessment in Software Engineering}, 2024.

\bibitem{doderleinPilotingCopilotCodex2022}
J.-B. D{\"o}derlein, M.~Acher, D.~E. Khelladi, B.~Combemale, Piloting
  {{Copilot}} and {{Codex}}: {{Hot Temperature}}, {{Cold Prompts}}, or {{Black
  Magic}}? (Jun. 2023).
\newblock \href {https://doi.org/10.2139/ssrn.4496380}
  {\path{doi:10.2139/ssrn.4496380}}.

\bibitem{Copilotexplorera}
P.~Thakkar,
  \href{https://thakkarparth007.github.io/copilot-explorer/posts/copilot-internals.html}{Copilot-explorer}
  (2024).
\newline\urlprefix\url{https://thakkarparth007.github.io/copilot-explorer/posts/copilot-internals.html}

\bibitem{li2023starcoder}
R.~Li, et~al., {{StarCoder}}: May the source be with you! (2023).
\newblock \href {http://arxiv.org/abs/2305.06161} {\path{arXiv:2305.06161}}.

\bibitem{lozhkovStarCoderStackV22024}
A.~Lozhkov, et~al., {{StarCoder}} 2 and {{The Stack}} v2: {{The Next
  Generation}} (Feb. 2024).
\newblock \href {http://arxiv.org/abs/2402.19173} {\path{arXiv:2402.19173}},
  \href {https://doi.org/10.48550/arXiv.2402.19173}
  {\path{doi:10.48550/arXiv.2402.19173}}.

\bibitem{netease-fuxiNetEaseFuXiEETQ2024}
{NetEase Fuxi Lab}, \href{https://github.com/NetEase-FuXi/EETQ}{{Easy \&
  Efficient Quantization for Transformers (EETQ)}} (May 2024).
\newline\urlprefix\url{https://github.com/NetEase-FuXi/EETQ}

\bibitem{dettmersTimDettmersBitsandbytes2024}
T.~Dettmers,
  \href{https://huggingface.co/docs/bitsandbytes/main/en/index}{{{TimDettmers}}/bitsandbytes},
  HuggingFace (May 2024).
\newline\urlprefix\url{https://huggingface.co/docs/bitsandbytes/main/en/index}

\bibitem{Ember2024}
{Ember}, Carbon intensity of electricity generation -- ember and energy
  institute (2024).

\bibitem{mozannarWhenShowSuggestion2024}
H.~Mozannar, G.~Bansal, A.~Fourney, E.~Horvitz, When to {{Show}} a
  {{Suggestion}}? {{Integrating Human Feedback}} in {{AI-Assisted
  Programming}}, AAAI Conference on Artificial Intelligence 38~(9) (2024)
  10137--10144.
\newblock \href {https://doi.org/10.1609/aaai.v38i9.28878}
  {\path{doi:10.1609/aaai.v38i9.28878}}.

\bibitem{barkeGroundedCopilotHow2022}
S.~Barke, M.~B. James, N.~Polikarpova, Grounded copilot: How programmers
  interact with code-generating models, Proc. ACM Program. Lang. 7~(OOPSLA1)
  (apr 2023).
\newblock \href {https://doi.org/10.1145/3586030} {\path{doi:10.1145/3586030}}.

\bibitem{MeasuringGitHubCopilot2024}
\href{https://cacm.acm.org/research/measuring-github-copilots-impact-on-productivity/}{Measuring
  {{GitHub Copilot}}'s {{Impact}} on {{Productivity}} -- {{Communications}} of
  the {{ACM}}} (Feb. 2024).
\newline\urlprefix\url{https://cacm.acm.org/research/measuring-github-copilots-impact-on-productivity/}

\bibitem{dimitropoulosEnergyProductivityImprovements2007}
J.~Dimitropoulos, Energy productivity improvements and the rebound effect:
  {{An}} overview of the state of knowledge, Energy Policy 35~(12) (2007)
  6354--6363.
\newblock \href {https://doi.org/10.1016/j.enpol.2007.07.028}
  {\path{doi:10.1016/j.enpol.2007.07.028}}.

\bibitem{borzunov2022petals}
A.~Borzunov, D.~Baranchuk, T.~Dettmers, M.~Riabinin, Y.~Belkada,
  A.~Chumachenko, P.~Samygin, C.~Raffel, Petals: {{Collaborative Inference}}
  and {{Fine-tuning}} of {{Large Models}}, in: 61st {{Annual Meeting}} of the
  {{Association}} for {{Computational Linguistics}} ({{Volume}} 3: {{System
  Demonstrations}}), Association for Computational Linguistics, Toronto,
  Canada, 2023, pp. 558--568.
\newblock \href {https://doi.org/10.18653/v1/2023.acl-demo.54}
  {\path{doi:10.18653/v1/2023.acl-demo.54}}.

\bibitem{simonUncoveringEnvironmentalImpact2023}
T.~Simon, P.~Rust, R.~Rouvoy, J.~Penhoat, Uncovering the {{Environmental
  Impact}} of {{Software Life Cycle}}, in: 2023 {{International Conference}} on
  {{ICT}} for {{Sustainability}} ({{ICT4S}}), 2023, pp. 176--187.
\newblock \href {https://doi.org/10.1109/ICT4S58814.2023.00026}
  {\path{doi:10.1109/ICT4S58814.2023.00026}}.

\bibitem{vaithilingamExpectationVsExperience2022}
P.~Vaithilingam, et~al., Expectation vs. {{Experience}}: {{Evaluating}} the
  {{Usability}} of {{Code Generation Tools Powered}} by {{Large Language
  Models}}, in: {{CHI Conference}} on {{Human Factors}} in {{Computing Systems
  Extended Abstracts}}, 2022, pp. 1--7.
\newblock \href {https://doi.org/10.1145/3491101.3519665}
  {\path{doi:10.1145/3491101.3519665}}.

\bibitem{papineniBLEUMethodAutomatic2002}
K.~Papineni, S.~Roukos, T.~Ward, W.-J. Zhu, {{BLEU}}: A method for automatic
  evaluation of machine translation, in: 40th {{Annual Meeting}} on
  {{Association}} for {{Computational Linguistics}}, {{ACL}} '02, Association
  for Computational Linguistics, 2002, pp. 311--318.
\newblock \href {https://doi.org/10.3115/1073083.1073135}
  {\path{doi:10.3115/1073083.1073135}}.

\bibitem{chenEvaluatingLargeLanguage2021}
M.~Chen, et~al., Evaluating {{Large Language Models Trained}} on {{Code}} (Jul.
  2021).
\newblock \href {http://arxiv.org/abs/2107.03374} {\path{arXiv:2107.03374}}.

\bibitem{pearceAsleepKeyboardAssessing2022}
H.~Pearce, et~al., Asleep at the {{Keyboard}}? {{Assessing}} the {{Security}}
  of {{GitHub Copilot}}'s {{Code Contributions}}, in: 2022 {{IEEE Symposium}}
  on {{Security}} and {{Privacy}} ({{SP}}), 2022, pp. 754--768.
\newblock \href {https://doi.org/10.1109/SP46214.2022.9833571}
  {\path{doi:10.1109/SP46214.2022.9833571}}.

\bibitem{sandovalSecurityImplicationsLarge2022}
G.~Sandoval, et~al., Lost at {{C}}: {{A User Study}} on the {{Security
  Implications}} of {{Large Language Model Code Assistants}}, in: 32nd {{USENIX
  Security Symposium}} ({{USENIX Security}} 23), 2023, pp. 2205--2222.

\bibitem{perryUsersWriteMore2022}
N.~Perry, et~al., Do {{Users Write More Insecure Code}} with {{AI
  Assistants}}?, in: 2023 {{ACM SIGSAC Conference}} on {{Computer}} and
  {{Communications Security}}, {{CCS}} '23, 2023, pp. 2785--2799.
\newblock \href {https://doi.org/10.1145/3576915.3623157}
  {\path{doi:10.1145/3576915.3623157}}.

\bibitem{jesseLargeLanguageModels2023}
K.~Jesse, et~al., Large {{Language Models}} and {{Simple}}, {{Stupid Bugs}},
  in: 2023 {{IEEE}}/{{ACM}} 20th {{International Conference}} on {{Mining
  Software Repositories}} ({{MSR}}), 2023, pp. 563--575.
\newblock \href {https://doi.org/10.1109/MSR59073.2023.00082}
  {\path{doi:10.1109/MSR59073.2023.00082}}.

\bibitem{yetistirenAssessingQualityGitHub2022}
B.~Yetistiren, et~al., Assessing the quality of {{GitHub}} copilot's code
  generation, in: 18th {{Int. Conf.}} on {{Predictive Models}} and {{Data
  Analytics}} in {{Software Engineering}}, 2022, pp. 62--71.
\newblock \href {https://doi.org/10.1145/3558489.3559072}
  {\path{doi:10.1145/3558489.3559072}}.

\bibitem{nguyenEmpiricalEvaluationGitHub2022}
N.~Nguyen, S.~Nadi, An empirical evaluation of {{GitHub}} copilot's code
  suggestions, in: 19th {{International Conference}} on {{Mining Software
  Repositories}}, 2022, pp. 1--5.
\newblock \href {https://doi.org/10.1145/3524842.3528470}
  {\path{doi:10.1145/3524842.3528470}}.

\bibitem{liu2024your}
J.~Liu, C.~S. Xia, Y.~Wang, L.~Zhang, Is your code generated by chatgpt really
  correct? rigorous evaluation of large language models for code generation,
  Advances in Neural Information Processing Systems 36 (2024).

\bibitem{austinProgramSynthesisLarge2021}
J.~Austin, et~al., Program {{Synthesis}} with {{Large Language Models}} (Aug.
  2021).
\newblock \href {http://arxiv.org/abs/2108.07732} {\path{arXiv:2108.07732}}.

\bibitem{yuCoderEvalBenchmarkPragmatic2023}
H.~Yu, et~al., {{CoderEval}}: {{A Benchmark}} of {{Pragmatic Code Generation}}
  with {{Generative Pre-trained Models}}, in: {{IEEE}}/{{ACM}} 46th
  {{International Conference}} on {{Software Engineering}}, ACM, 2024, pp.
  1--12.
\newblock \href {https://doi.org/10.1145/3597503.3623316}
  {\path{doi:10.1145/3597503.3623316}}.

\bibitem{hendrycksMeasuringCodingChallenge2021}
D.~Hendrycks, et~al., Measuring {{Coding Challenge Competence With APPS}},
  Neural Information Processing Systems Track on Datasets and Benchmarks 1
  (Dec. 2021).

\bibitem{luCodeXGLUEMachineLearning2021}
S.~Lu, et~al., {{CodeXGLUE}}: {{A Machine Learning Benchmark Dataset}} for
  {{Code Understanding}} and {{Generation}}, Neural Information Processing
  Systems Track on Datasets and Benchmarks 1 (Dec. 2021).

\bibitem{wangReCodeRobustnessEvaluation2022}
S.~Wang, et~al., {{ReCode}}: {{Robustness Evaluation}} of {{Code Generation
  Models}}, in: 61st {{Annual Meeting}} of the {{Association}} for
  {{Computational Linguistics}} ({{Volume}} 1: {{Long Papers}}), Association
  for Computational Linguistics, Toronto, Canada, 2023, pp. 13818--13843.
\newblock \href {https://doi.org/10.18653/v1/2023.acl-long.773}
  {\path{doi:10.18653/v1/2023.acl-long.773}}.

\bibitem{xuSystematicEvaluationLarge2022}
F.~F. Xu, et~al., A systematic evaluation of large language models of code, in:
  6th {{International Symposium}} on {{Machine Programming}}, {{MAPS}} 2022,
  2022, pp. 1--10.
\newblock \href {https://doi.org/10.1145/3520312.3534862}
  {\path{doi:10.1145/3520312.3534862}}.

\bibitem{luccioniCountingCarbonSurvey2023}
A.~S. Luccioni, A.~{Hernandez-Garcia}, Counting {{Carbon}}: {{A Survey}} of
  {{Factors Influencing}} the {{Emissions}} of {{Machine Learning}} (Feb.
  2023).
\newblock \href {http://arxiv.org/abs/2302.08476} {\path{arXiv:2302.08476}}.

\bibitem{selvanEquityAccessCase2024}
R.~Selvan, B.~Pepin, C.~Igel, G.~Samuel, E.~B. Dam, Equity through {{Access}}:
  {{A Case}} for {{Small-scale Deep Learning}} (Mar. 2024).
\newblock \href {http://arxiv.org/abs/2403.12562} {\path{arXiv:2403.12562}},
  \href {https://doi.org/10.48550/arXiv.2403.12562}
  {\path{doi:10.48550/arXiv.2403.12562}}.

\bibitem{chakravartyDeepLearningModels2024}
A.~Chakravarty, Deep {{Learning Models}} in {{Speech Recognition}}: {{Measuring
  GPU Energy Consumption}}, {{Impact}} of {{Noise}} and {{Model Quantization}}
  for {{Edge Deployment}} (May 2024).
\newblock \href {http://arxiv.org/abs/2405.01004} {\path{arXiv:2405.01004}}.

\end{thebibliography}

\end{document}